\documentclass[12pt,journal,compsoc]{IEEEtran}
\ifCLASSOPTIONcompsoc
\else
\fi
\ifCLASSINFOpdf
\else
\fi
\usepackage[dvips]{graphicx}
\usepackage{subfigure}

\begin{document}
\title{Variance Based Algorithm for Grouped-Subcarrier Allocation in OFDMA Wireless Systems} 
\author{Noureddine {Hamdi}, ~\IEEEmembership{Member,~IEEE,}}

\maketitle

\begin{abstract}
In this paper, a reduced complexity algorithm is proposed for grouped-subcarriers and power allocation in the downlink of OFDMA packet access wireless systems.
The available subcarriers for data communication are grouped into partitions (groups) where each group is defined as a subchannel. The scheduler located at the base station allocates subchannels to users based on the variance of subchannel gains. The proposed algorithm for group allocation is a two-step algorithm that allocates groups to users based on the descending order of their variances to resolve the conflicting selection problem, followed by a step of fairness proportionality enhancement. To reduce the feedback burden and the complexity of the power allocation algorithm, each user feeds back the CSI on each group if the variance of gains of subcarriers inside it is less than a predefined threshold. To Show the performance of the proposed scheme, a selection of simulation results is presented.
\end{abstract}

\keywords{
OFDMA; grouped-subcarrier; Resource allocation; fairness; performance analysis.
}
\section{Introduction}
Optimal and suboptimal algorithms for radio resource management in wireless high speed networks based on orthogonal frequency division multiplexing (OFDM) are widely studied in the literature. Several systems are considered in these studies: single or multiple antennas at the transmitter and at the receiver. The complexity of resource allocation algorithms can be reduced in the case of downlink environment of OFDM access (OFDMA) systems by dividing the available subcarriers into groups (partitions) based on the channel coherence bandwidth and users compete for these groups. This system generalizes the independent subcarrier per subcarrier allocation. Recent published studies have considered this system model as \cite{Kim2005} \cite{Alen2003} \cite{Xiaowen03} \cite{Han2007} and references therein. The groups are composed of continuous or non-continuous subcarriers, based on how the channel fading is fast or slow as described in \cite{Ibing2007}.

Multiuser diversity (MUD), if used optimally, enhances the system sum capacity by mitigating the fading effect, while providing some quality of service (QoS) to users. For a given user QoS requirement, optimal algorithms have to optimize one between two major goals: (i) maximize the data rate under power constraints \cite{Alen2003} or (ii) minimize the overall transmit power under data rate constraints \cite{Chen2004}. 
Some papers considered the first optimization problem for the radio resource management of OFDMA systems.
Therefore, an adaptive subcarrier allocation algorithm has been proposed in \cite{Xiaowen03}, where the available subcarriers are divided into blocks (groups) and the algorithm assigns blocks to each user according to its required data rate and BER constraint. This approach 
is a two-step method that firstly, adopted an adaptive block allocation to increase the system capacity by using channel state information (CSI) of all users and assigning a subset of blocks with the highest average channel gains to the corresponding user. Secondly, an iterative improvement procedure is employed to minimize the total required transmit power while satisfying multiuser data rates and BER requirements. 
This algorithm enhances the throughput and the power efficiency but doesn't provide a capacity close to the optimal.

To increase the system capacity a decentralized subcarrier allocation algorithm \cite{Alen2003} provided an interesting solution. All users divide all the subcarriers into a number of partitions (groups) in parallel and each user selects the partition with the highest average channel gain independently. Since each user attempts to select the partition with the highest average channel gain, more than one user may conflict in the selection of a partition, if it is the best one to them simultaneously. The important contribution of this algorithm is to resolve the conflict selection. 
An iterative algorithm is used for partition with conflicting selection by using a usage value of each partition for each user. This iterative step enhances the system performance but increases significantly the algorithmic complexity of the system.

A solution to this problem has been provided in \cite{Kim2005}, where an adaptive grouped-subcarrier allocation algorithm is proposed for increasing the system broadcast capacity of OFDM systems and reducing their complexity. Instead of the usage value, the known average channel gain of each group is used to resolve the conflict problem among users. The first step of the allocation algorithm is followed by a swap procedure used to enhance the system capacity. This swap procedure is its still drawback. Therefore, it enhances the system capacity but for the cost of a significant increase of the complexity.

As a solution to this problem, we propose an algorithm based on the variance of the transmissible rates on user groups. 
The groups without conflict selection are assigned first. Then, the scheduler computes, for each active user, the variance of transmissible rates on all the reported groups (the variance of the reported gains). These variances are then sorted in a descending order. To optimally share the available groups, the proposed algorithm allows first, the user that has the max variance to select his best group. The assigned group is removed from the set of the available groups for the allocation. The same processing is done for the set of the remainder groups. If a user reaches his requested rate, he would be removed from the set of users allowed to compete for group allocation. After this first step, the subset of the remainder unassigned groups is allocated adaptively to users. Accordingly, while providing a system capacity close to that obtained in \cite{Kim2005}, the complexity is reduced efficiently. This algorithm would be analyzed in section \ref{Prop}.

The remainder of this paper is organized as follows: While section \ref{Sys} presents the system model, section \ref{Prop} analyses the operation mode of the proposed scheduling scheme and section \ref{Sim} presents a selection of simulation results and performance analysis.

\section{System model and operating mode}
\label{Sys}
\begin{figure}[htbp]
\centerline{
\includegraphics[scale=0.5]{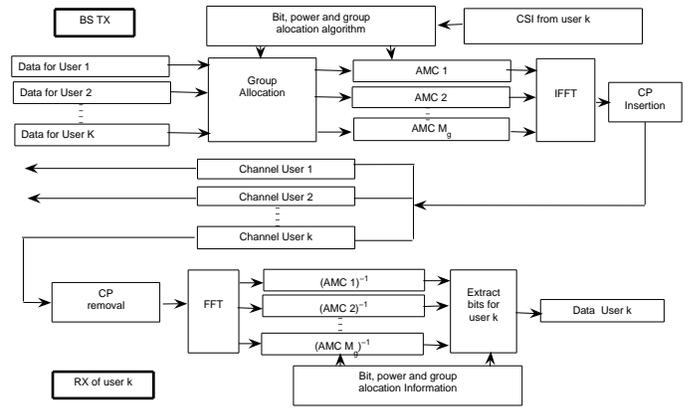}
}
\caption{OFDMA TX-RX system chain}
\label{SysM}
\end{figure}

The system model is similar to that presented in \cite{Alen2003} and shown in Figure \ref{SysM}.
We limit our consideration to a simplified system model. The system consists of a single base station (BS) that serves $K$ users, each employing OFDM transmission over a subset of the available M subcarriers. Power and groups of subcarriers are allocated to users according to their CSI and their user QoS constraints.

At the transmitter, by using the CSI from the $K$ users, the allocation algorithm is applied to assign different groups of subcarriers to different users. The number of bits to be transmitted on subcarriers of each group is also determined in the process. This information is used to configure the adaptive modulators and the input data should be modulated accordingly. The complex symbols at the output of the adaptive modulators and coders (AMC) would then be transformed into time-domain samples by the IFFT block. Next, a cyclic prefix (CP) is inserted to each time-domain sample for inter symbol interference (ISI) cancellation and to ensure orthogonality in multipath environments. The transmit signal is then sent through different frequency selective fading channels to different users.
At the receiver, the CP is removed after ISI cancellation and the time samples of the $k^{th}$ user are transformed by the FFT block into modulator symbols. The bit allocation information is used to configure the adaptive demodulators and decoders $(AMC)^{-1}$, while the group allocation information is used to extract the demodulated bits from the subcarriers assigned to the $k^{th}$ user.

The proposed scheme assigns non-continuous subcarriers which spreads the user symbols over a wider frequency range. Thus, frequency diversity is exploited and is considered for high-mobility scenarios as proposed in \cite{Ibing2007} and in \cite{Giannakis2003}. The goal of this work is to enhance the system sum capacity by exploiting maximum diversity gain. If we have $M=M_gN_g$ subcarriers distributed into $M_g$ groups of $N_g$ subcarriers, the $m^{th}$ group has the subcarriers $\left\lbrace m, m+M_g,...,\left( M_g-1\right) N_g+m\right\rbrace $. This simple method provides an interesting diversity gain which is analyzed in \cite{Giannakis2003}.

The allocation algorithm operates in a discrete-time fashion. Time is divided between short guard period followed by longer data burst. The guard period is used to perform a series of operations, including calculus of variance of transmissible rates and user QoS comparisons to reach the appropriate decision. 
Each user reports to the BS each group having a variance of the $N_{g}$ subcarriers inside it, to be less than a predefined threshold. Then, we have up to $M_g=\frac {M} {N_{g}}$ users that are allowed to receive (for downlink) packets throughout the subsequent data burst time.
The considered model for the channel is a multipath channel that has $L_p$ paths, where each path has a complex amplitude $h_l(t$) and a delay $\tau_l$. The time and frequency channel responses are respectively:
\begin{equation}
\label {ChR}
h(t,\tau)=\sum_{l=0}^{L_p-1}h_l(t)\delta(\tau-\tau_l),\; H(t,f)=\sum_{l=0}^{L_p-1}h_l(t)e^{-j2\pi f \tau_l}
\end{equation}
The sampled channel frequency response (eq. \ref{ChR}) of user $k$ on subcarrier $m$ in the OFDM symbol $n$ is:
\begin{equation}
\label {FR}
H_{k,m}=H(nT_s,k\Delta f)=\sum_{l=0}^{L_p-1}h_l(nT_s)e^{-j2\pi m\Delta f \tau_l}
\end{equation}
where $\Delta f$ and $T_s$ are respectively the subcarrier spacing and the OFDM symbol period.
After removing the CP and pilots and applying FFT, the sampled channel response of user $k$ is denoted $H_k$. The $k^{th}$ user receives a discrete baseband OFDM symbol from the serving BS:
\begin{equation}
\label {RS}
y_k = \sqrt{\rho}H_k x_k+n_k 
\label{eq1}
\end{equation}

Where $\rho=\frac{P_k}{N_0\Delta f}$ denotes the power to noise ratio, $x_k$ and $y_k$ are $M\times1$ vectors denoting the transmitted and the received signal, and $n_k$ is the noise bloc assumed to be a circular symmetric complex Gaussian random vector with distribution $CN\left(0, I_M\right)$, where $I_M$ is M-dimension identity matrix. The bandwidh of each subcarrier is assumed to be narrower than the coherent bandwidth of the channel and analog (Continuous) rate is considered. Then, the channel is viewed as a Gaussian channel. If $M$ subcarriers are used for data transmission, the transmissible rate of subcarrier m, when one user is considered, is: $C_{k,m}=log_2\left(1+ \frac{\gamma_{k,m}}{\Gamma}\right)$,where $\gamma_{k,m} = \rho_{k,m}\left|H_{k,m}\right|^2$ is the SNR on the $m^{th}$ subcarrier, and $H_{k,m}$ is the user subchannel gain. $\Gamma$ is the SNR gap used to satisfy the BER constraint. If a square quadrature amplitude modulation $QAM$ with N-symbol-Gray-mapping is the base band modulation technique, the SNR gap for a given BER is approximated by: $\Gamma\simeq-\frac{\ln(5BER)}{1.6}$ if $N\geq4$ (deduced from \cite{Chung01}, AWGN Channel).

\section{Variance based resource allocation}
\label{Prop}
The first step of the proposed scheme is done by the user mobile station (MS) which would proceed some operations: measure the channel gains of subcarriers of each group, and then report to the BS each group inside it subcarriers with a variance lower than a predefined threshold as: $V_{k,m} \leq\epsilon {\bar\gamma_{k,m}}^2$, where $V_{k,m}$ an ${\bar\gamma_{k,m}}$ are the variance and the average of the subcarriers gains inside group $m$ of user $k$. $\epsilon$ (typical value $0.5$) is used to define the groups whose subcarriers have roughly equal gains. The second step starts by computing the variance of the reported subchannel gains. The goal of the scheduling algorithm is assigning the set of subchannels (groups) to meet the highest system transmissible rate under proportional fairness constraints (PFC). After examination of the reported CSI (subchannel gains) on all groups, groups without conflict selection among users are assigned fist, since each is requested by a single user. The remainder unassigned groups are requested by at least two users. Then, the group allocation algorithm has to select for each group a user that maximizes the objective function under the system constraints. To explain the proposed idea, we consider an example, where two users share 4 groups: the first has subchannel gains of groups with a low variance, and the second has a high variance. Consequently, if each user compete for two groups among the four available, the system capacity will be greater, if the user with high variance selects first. This can be better seen, if numerical samples are considered as presented in the following:

\begin{tabular}{l|c|c|c|c}
& \textbf{G1} & \textbf{G2} & \textbf{G3} & \textbf{G4} \\
\hline
\textbf{Transmissible rates of User 1} & 90 & 60 & 20 & 10 \\
\hline
\textbf{Transmissible rates of User 2} & 100 & 90 & 70 & 70 \\
\end{tabular}
\\
To simplify the analysis, we assume that inside each group one subcarrier. According to the proposed algorithm, we allow user with high variance to select first. The variances of user 1 and user 2 are respectively $V_1 = 1367$ and $V_2 = 225$. If we select for each group the best user under the constraint that each user have to select up to two groups, the system receives $R_{best}= 100 +90 +20 +10 = 220$. If we apply the variance based algorithm, the system receives $R_{var} = 90 + 60 +70 + 70 = 290$. 
Thus, as shown in this simple example, the proposed variance based algorithm can provide better performance. This simple example can be generalized to systems with slow power control where variances are widely spread, and this agrees, with the high-mobility scenarios.

Then the system capacity increases if, the algorithm starts by assigning to the user with the highest variance, his best group of subcarriers. This group is then removed from the set of groups at the scheduling disposal. Then, the scheduler has to process with the same way the set of the remainder unassigned groups. In a first step, according to the PFC, if a user has reached his requested number of groups, he would be removed from the set of users allowed to compete for groups. In the second step, if there are groups not yet assigned, the scheduler has to enhance the system throughput and to meet the PFC in throughputs. Then, for each unassigned group, the scheduler assigns it to the user that has the lowest average rate weighted by his proportional coefficient among users that have the $L$ best transmissible rates. Then, the throughput proportionality is obtained through two steps as:

\begin{enumerate}
\item first PFC is considered for the number of assigned groups as: $M_{g,1}:M_{g,2}:..M_{g,K}=\alpha_1: \alpha_2:..:\alpha_K$, where $M_{gk}$ is the number of groups assigned to user $k$.
\item After the second step, we meet the PFC in throughputs as: $R_1:R_2:..:R_K=\alpha_1: \alpha_2:..:\alpha_K$.
\end{enumerate}
In the following, the key steps of this scheduling algorithm are given.\\
\textbf{Subchannel allocation}
\begin{enumerate}
\item \textit{Initialization} \\
\textit{Equal power is allocated to groups}; \\
U = \{ active users (MS) that has reported gains \}; \\ 
$U_s=\emptyset$, ({Users that has obtained their requested groups});\\
$S=$ \{ subchannels (groups) available for assignment \}; \\
$S_R=\emptyset$, ({the set of assigned groups});\\
$G_S=$ \{ subchannel gains on the elements of S\}; \\
$S_k =$ \{groups reported by user k\};\\
$V=\left\lbrace v_k=var(S_k), k=1..K\right\rbrace $; $A = \left\lbrace \alpha_k, k=1..K\right\rbrace $; \\
$M_A= \left\lbrace M_k = \lfloor\alpha_k M_g\rfloor, k=1..K \right\rbrace $ ;\\
$MaxIt = M_g$, $L = \frac{K}{4}$, (For example); 
\item \textit{First step}\\
i = 1, \\
\underline{{While}} ($i\leq MaxIt$), \\
$k_s = \arg \max_{v_k \in V} v_k$ ; $m_{k_s} = \arg \max_{{g_k}\in S_k} g_k$; \\
\underline{{If}} $M_A(k_s) \geq 1 $ ;\\
Assign group $m_{k_s}$ to user $k_s$ ; \\
$M_A(k_s)= M_A(k_s)-1$ ;
$S_k = S_k\setminus\{m_{k_s}\}$; $S=S\setminus\{m_{k_s}\}$;\\ $S_R = S_R\cup \{m_{k_s}\}$, \underline{{Else}}, $U = U\setminus\{k_s\}$;$U_s=U_s\cup\{k_s\}$\\
\underline{{EndIf}}, 
$i=i+1$; update $V$; 
\underline{{EndWhile}} 
\item \textit{Second step}\\
$U=U\cup U_s$
For all $m\in S\setminus S_R$ do \\
Sort in descending order transmissible rates \\
Assign group $m$ to user with $\min \frac{R_k}{\alpha_k}$ between those owning the $L$ best subchannel gains (the first $L$ elements of the sorted rates).
\end{enumerate}
\textbf{Computational complexity}\\
Resource allocation complexity is a main criterion in the design of suboptimal algorithms. This analysis is processed in the case of a system with $K$ active users and $M_g$ available subchannels. The initialization step requires the computation of $K$ variances and sorting them in a descending order, which requires $K M_g$ multiplication, $K (M_g-1)$ addition and $K \log(K)$ operation for the sorting. Thus, the complexity is $O(K M_g)$.
The step 1 involves a loop of $Itmax$ iterations where variances of users subchannels' gains in a descending order followed by updating these variances. This requires $O(Itmax M_g \log K)$ operations.
In Step 3, each unassigned subchannel in step 2 is allocated to the worst case user among the $L$ best users. The number of still available subchannels is $M_g^*=\bar S$, where $\bar{(.)}$ is the cardinality of $(.)$. This step requires $O(K L\log(M_g^*)$ operations.

\textbf{Power allocation } 
The OFDM transmitter allocates equal transmit power to the subchannels under total power $P_t$ constraint: $\sum_{k=1}^K P_k\leq P_t$. The subchannel allocation steps allocate to the $k^{th}$ user $M_{g,k}$ subchannels, where $k=1..K$. Then, each user receives $P_k=P_t \frac{M_{g,k}}{M_g}$ which is shared equally by the assigned subchannels. Since the $N_g$ subcarriers inside a given subchannel (group) have roughly equal gain, the power assigned to the considered group is distributed equally to them. Then, each subcarrier receives $p_{k,m}=\frac{P_k}{M_{g,k}N_g}$. 

\section {Simulation Results}
\label{Sim}
\begin{figure}[htbp]
\centerline{
\includegraphics[width=9cm]{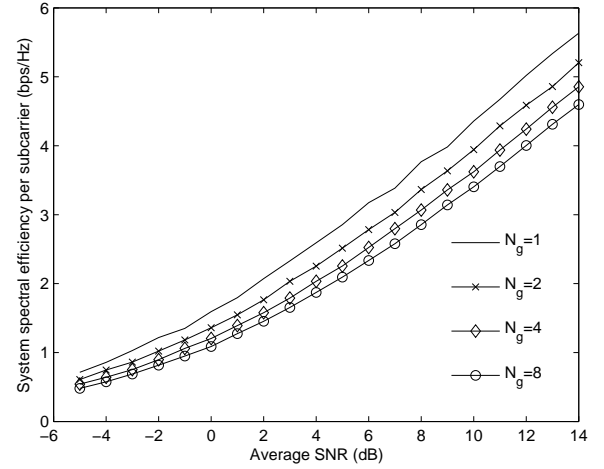}
}
\caption{System average throughput per subcarrier, different number of subcarriers per group $N_g$}
\label{SThPrNg}
\end{figure}

\begin{figure}[htbp]
\centerline{
\includegraphics[width=9cm]{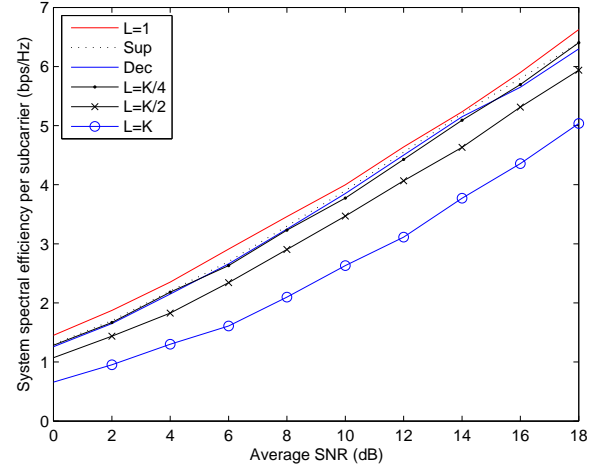}
}
\caption{System throughput per subcarrier, proposed (different $L$), decentralized \cite{Alen2003} "Dec", comparative superiority algorithms \cite{Kim2005} "Sup"}
\label{SThPrL}
\end{figure}

In this section, simulation results are analyzed to show the performance of the proposed algorithm compared to comparative superiority \cite{Kim2005} and decentralized \cite{Alen2003} algorithms used as references for the performance comparison.
The system considered for simulation is the downlink of a single cell that uses $M = 128$ subcarriers for data communications and serves $K = 8$ mobile users. The considered SNR gap is $\Gamma=1$ and the threshold $\epsilon = 0.5$. According to the assumption of power control and Rayleigh fading, the instantaneous SNR of each subcarrier is modeled by an i.i.d. exponential distribution. 
In the beginning of this analysis, the number of subcarriers per group $N_g$ is considered to be variable ($N_g =1,2,4,8$), the parameter $L = K/4$, the simulation window $T = 200$ time slots, and the proportionality coefficients are $\alpha = [2;1;3;1;2;2;4;4]/19$. Figure \ref{SThPrNg} shows that, if the number of subcarriers per group ($N_g$) increases the system diversity gain decreases and consequently, the system throughput would decrease.
To compare the variance based algorithm performance, the system throughput per subcarrier vs. the average SNR(dB) is illustrated. Figure \ref{SThPrL} shows the dependence of the system throughput to the parameter $L$. As $L$ increases, the proportionality fairness performance is enhanced, as will be shown below, for the price of a loss in system throughput. The illustrations compare also, our algorithm to the comparative superiority (Sup) \cite{Kim2005} and the decentralized algorithm (Dec) \cite{Alen2003}. The proposed scheme when $L$ is kept to be $\frac{K}{4}$ provides the same throughputs as \cite{Kim2005}. The advantage of the proposed scheme is the reduction of the complexity. 

\begin{figure}[htbp]
\centerline{
\includegraphics[width=11cm]{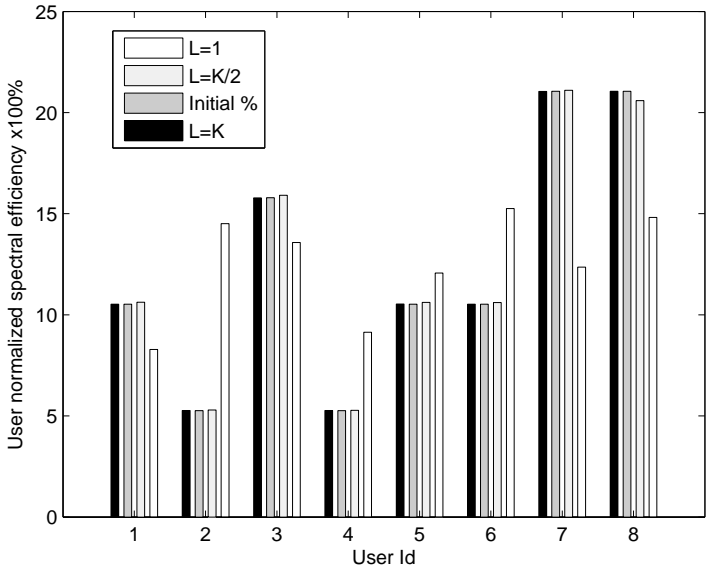}
}
\caption{User normalized throughput for different $L$, Initial \% are the proportional coefficients $\alpha_k$.}
\label{UTh}
\end{figure}

\begin{figure}[htbp]
\centerline{
\includegraphics[width=9cm]{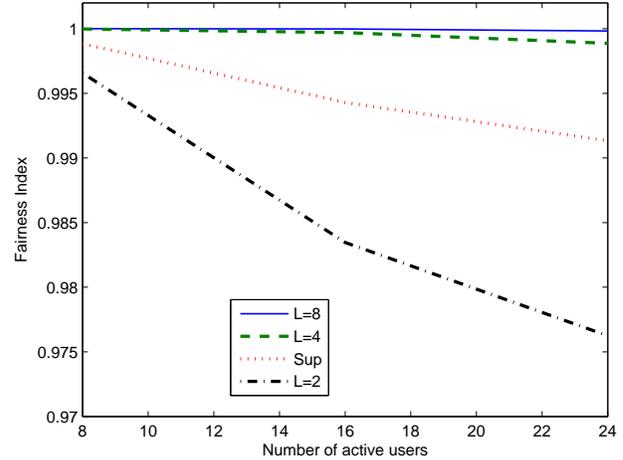}
}
\caption{Jain's fairness index}
\label{Fair}
\end{figure}

With the same simulation parameters, figure \ref{UTh} shows the different normalized throughputs per user. As illustrated in this figure, as $L$ taken near $K$ ($L = K$, $L = \frac{K}{2}$ ), there is a strict rate proportionality. But if $L=1$, for each unassigned group in the first step, the user with the best transmissible rate is selected without consideration of the proportionality constraint. 
This study is completed by comparing the proportional fairness degree of the proposed scheme when $L = 2$, $L = 4$ or $L = 8$. Figure \ref{Fair}, illustrates the modified Jain's fair index \cite{Jain1991} used as a metric for the proportional fairness comparison of weighted throughputs vs. the number of active users in the cell. This modified index is expressed as: $F_I = \frac{\left(\sum_{k=1}^{K}\frac{R_k}{\alpha_k}\right)^2}{K\sum_{k=1}^{K}\left(\frac{R_k}{\alpha_k}\right)^2}$
As shown in this figure, the Jain proportional fairness indexes for different values of $L$ are about $0.99$ which is quasi-optimal (optimal $=1$). The case $L=1$ is not considered in this comparison because, it doesn't agree with the PFC. If $L$ increases the fairness index increases. The comparative superiority algorithm (Sup) is used as a benchmark. The fairness index of the proposed algorithm is better than that of "Sup" if $L \geq 4$ for $K=8$ to $24$.
\section{Conclusion}
\label{Conc}
A variance based adaptive grouped subcarrier allocation algorithm is proposed for the enhancement of the system sum capacity in multiuser OFDMA packet access systems. Compared to other algorithms, the proposed allocation algorithm reduces efficiently the complexity while adaptively enhance the system capacity under proportional fairness constraints.
\nocite{}
\bibliographystyle{unsrt}
\bibliography{OFDMAgroup3}

\begin{thebibliography}{1}

\bibitem{Kim2005}
N.~Womack~B.F. Youngok, K.~Haewoon.
\newblock An adaptive grouped-subcarrier allocation algorithm using comparative
  superiority.
\newblock IEEE Military Communications Conference, IEEE MILCOM 2005.

\bibitem{Alen2003}
A.~S. Chin~F. Alen, T. C.~Madhukumar.
\newblock Capacity enhancement of a multi-user ofdm system using dynamic
  frequency allocation.
\newblock {\em IEEE Trans. Broadcast.}, 49:344--353, 2003.

\bibitem{Xiaowen03}
Z.~Xiaowen L.~Jinkang.
\newblock An adaptive subcarrier allocation algorithm for multiuser ofdm
  system.
\newblock Proc. IEEE Vehicular Technology, VTC 2003 Fall.

\bibitem{Han2007}
Y.~Han, S.~Han.
\newblock A competitive fair subchannel allocation for ofdma system using an
  auction algorithm.
\newblock Proc. IEEE Veh. Technol. Conf. Baltimore, Maryland, USA, --VTC
  2007-Fall, 2007.

\bibitem{Ibing2007}
V.~Ibing, A.~Jungnickel.
\newblock On hardware implementation of multiuser multiplexing for sc-fdma.
\newblock Proc. IEEE Veh. Technol. Conf. VTC 2007-Fall.

\bibitem{Chen2004}
J.~Li~C. Chen, Y.~Chen.
\newblock A fast suboptimal subcarrier, bit, and power allocation algorithm for
  multiuser ofdm-based systems.
\newblock IEEE International Conf. on Communic., ICC 2004.

\bibitem{Giannakis2003}
Y.~Giannakis~G.B. Liu, Z.~Xin.
\newblock Linear constellation precoding for ofdm with maximum multipath
  diversity and coding gains.
\newblock {\em IEEE Trans. on Comm., TC--}, 51:416--427, 2003.

\bibitem{Chung01}
A.J. Chung, S.T.~Goldsmith.
\newblock Degrees of freedom in adaptive modulation: A unified view.
\newblock {\em IEEE Trans. on Comm., TC--}, 49:1561 -- 1571, 2001.

\bibitem{Jain1991}
R.~Jain.
\newblock The art of computer systems performance analysis.
\newblock John Wiley and Sons, 1991.

\end{thebibliography}
\end{document}